\documentclass[aps,prl,reprint,showpacs,superscriptaddress]{revtex4-1}
\usepackage{epsfig,graphicx,amsmath,amssymb}

\begin{document}

\title{Dispersion of particles in an infinite-horizon Lorentz gas}

\author{Lior Zarfaty}
\affiliation{Department of Physics, Institute of Nanotechnology and Advanced Materials, Bar-Ilan University, Ramat-Gan 52900, Israel}

\author{Alexander Peletskyi}
\affiliation{Institute of Physics, University of Augsburg, Universit\"atsstrasse 1, D-86135 Augsburg Germany}
\affiliation{Sumy State University, Rimsky-Korsakov Street 2, 40007 Sumy, Ukraine}

\author{Itzhak Fouxon}
\affiliation{Department of Physics, Institute of Nanotechnology and Advanced Materials, Bar-Ilan University, Ramat-Gan 52900, Israel}

\author{Sergey Denisov}
\affiliation{Institute of Physics, University of Augsburg, Universit\"atsstrasse 1, D-86135, Augsburg Germany}
\affiliation{Department of Applied Mathematics, Lobachevsky State University of Nizhny Novgorod, Gagarina Avenue 23, Nizhny Novgorod, 603950, Russia}

\author{Eli Barkai}
\affiliation{Department of Physics, Institute of Nanotechnology and Advanced Materials, Bar-Ilan University, Ramat-Gan 52900, Israel}

\begin{abstract}

We consider a two-dimensional Lorentz gas with infinite horizon. This paradigmatic model consists of pointlike particles undergoing elastic collisions with fixed scatterers arranged on a periodic lattice. It was rigorously shown that when $t\to\infty$, the distribution of particles is Gaussian. However, the convergence to this limit is ultraslow, hence it is practically unattainable. Here we obtain an analytical solution for the Lorentz gas' kinetics on physically relevant timescales, and find that the density in its far tails decays as a universal power law of exponent $-3$. We also show that the arrangement of scatterers is imprinted in the shape of the distribution.

\end{abstract}

\maketitle

The Lorentz gas (LG) is a classical model of transport \cite{Bunimovich,BSC}, in which a pointlike particle moves at a constant speed, while undergoing elastic collisions with fixed scatterers \cite{Bunimovich,BSC,Sanders,Courbage,Artuso,channel,Balint,Chernov,Dav,Ott,MZ,Fran,Burioni,Bianchi,Armstead,Szas,dett}. When the free paths in this model are unbounded, it is termed the infinite-horizon LG (see Ref.~\cite{Carl} for a review). Originally suggested as a description for the movement of electrons through a conductor, it is one of the first deterministic models to show a superdiffusive behavior, and as such it plays an important role in studying diffusion phenomena. Beautiful properties of the transport and ergodicity of the infinite-horizon LG were thoroughly investigated by mathematicians. For the two-dimensional case, Bleher \cite{Bleher} showed that $\lim_{t\to\infty}[\boldsymbol{r}(t)-\boldsymbol{r}(0)]/\sqrt{t\ln(t)}$ is a Gaussian variable, where $\boldsymbol{r}(t)$ is the particle's position at time $t$. However, this limit theorem hides the physically observable nature of the process \cite{Bleher,Cristadoro}, as it is valid only when $\ln[\ln(N)]/\ln(N)=\epsilon\ll 1$, where $N$ is the number of collisions. Thus, even if $N$ is very large, this condition cannot be satisfied (e.g., $\epsilon=0.01\Rightarrow N\approx 10^{281}$). In addition, as noted by Dettmann \cite{Carl}, the variance of this Gaussian limiting law is half of the mean-square displacement, which suggests that the far tails of the expanding packet deviate from a Gaussian.

\begin{figure*}
	\includegraphics[width=0.9\textwidth]{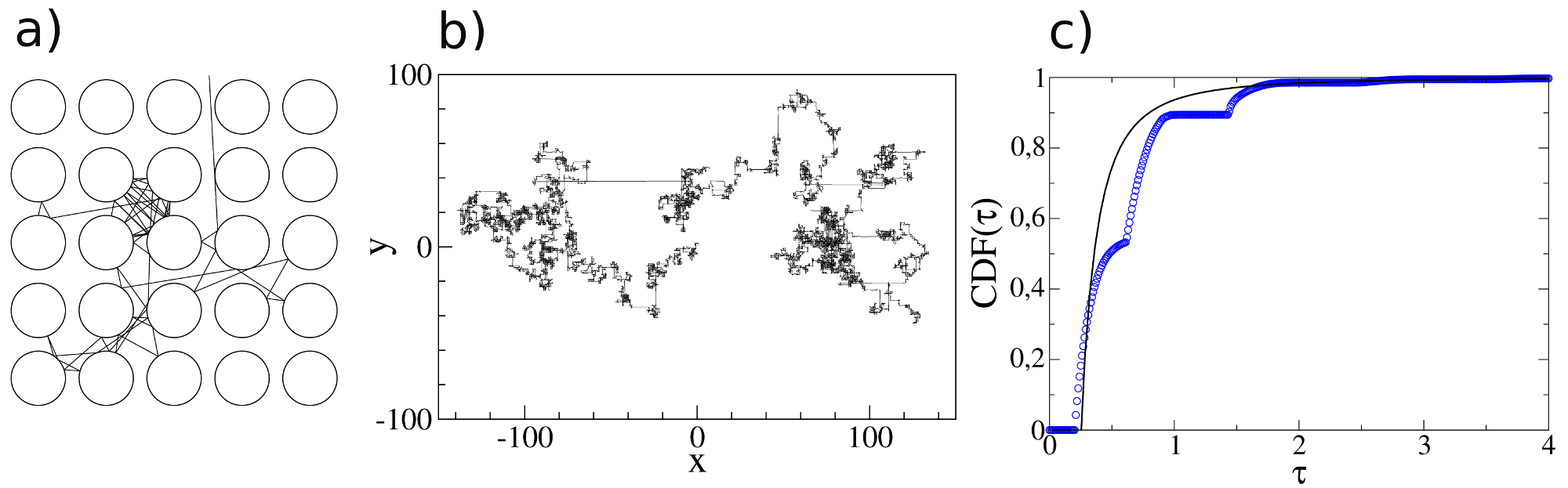}
	\caption{(a) In the two-dimensional Lorentz gas, a particle is moving with a constant speed, elastically colliding with circular scatterers residing on a square lattice. (b) Due to the infinite corridors, the particles exhibit long flights along the axes. (c) These power-law distributed times of travel are responsible to the power-law decay of their respective probability density function, Eq.~(\ref{eq01}), resulting in a cumulative distribution function (CDF) of the form $\text{CDF}(\tau)=\int_0^{\tau}\text{d}\tau'\psi(\tau') \simeq 1-\tau_0^2/\tau^2$ (solid line). The stairlike structure of the CDF (circles) originates from the discrete nature of the lattice of scatterers. We used $R=0.4$ as the radius of the scatterers, with lattice constant of $a=1$. The speed $V$ was chosen to be one.}
	\label{fig1}
\end{figure*}

In this Rapid Communication, we present a theory which captures the kinetics of the packet's density on physically attainable time scales and describes correctly its tails. By employing the L\'evy walk (LW) formalism, we go beyond the Gaussian description and use what we call a Lambert scaling approach, which converges already when $N \sim 10^4$. We show that in the far tail, the density decays as a universal power law, which is valid for transport models with an infinite horizon. Without loss of generality, we focus on circular scatterers of radius $1/\sqrt{8}<R<1/2$ which are placed on an infinite square lattice of constant $a=1$ [see Fig.~\ref{fig1}(a)]. We illustrate our analytical findings with numerical simulations.

{\em Fat-tailed traveling times.}
In an infinite horizon LG, a particle's trajectory exhibits intermittency. Namely, the particle undergoes epochs of diffusivelike behavior with many random reorientations, after which it follows an almost ballistic path within the endless corridors [see Fig.~\ref{fig1}(b)]. This behavior leads to long travel times $\{\tau_n\}$ between collision events, for which the probability density function (PDF) of $\tau$ follows a fat-tailed law \cite{Bou85,Zacharel}, such that its variance diverges just marginally,
\begin{equation}
\label{eq01}
	\lim_{\tau\to\infty}\tau^3\psi(\tau)=\tau_0^2 .
\end{equation}
Importantly, Eq.~(\ref{eq01}) is valid for spatial dimensions $d<6$ \cite{Carl}. The displacement of the particle is $\boldsymbol{r}(t)-\boldsymbol{r}(0) = \sum_{n=1}^{N} \boldsymbol{v}_{n-1}\tau_n + \boldsymbol{v}_{N}\tau^*$. Here, $N$ is the random number of collisions until time $t$, $\tau_n$ is the walking time of the $n$th travel epoch, $\boldsymbol{v}_n$ with $n\ge 1$ is the velocity just after the $n$th collision, $\boldsymbol{v}_0$ and $\boldsymbol{r}(0)$ are the initial velocity and displacement which are both randomly chosen, and the last traveling event is of duration $\tau^{*}=t-\sum_{n=1}^N\tau_n$. During this process the particle's speed is fixed due to the collisions' elasticity, and we choose $V=|\boldsymbol{v}_n|=1$. Our key assumption is that the LG model, being a chaotic system, can be described as a renewal process. This means we assume no correlation between two adjacent velocities.

\begin{figure*}
	\includegraphics[width=0.80\textwidth]{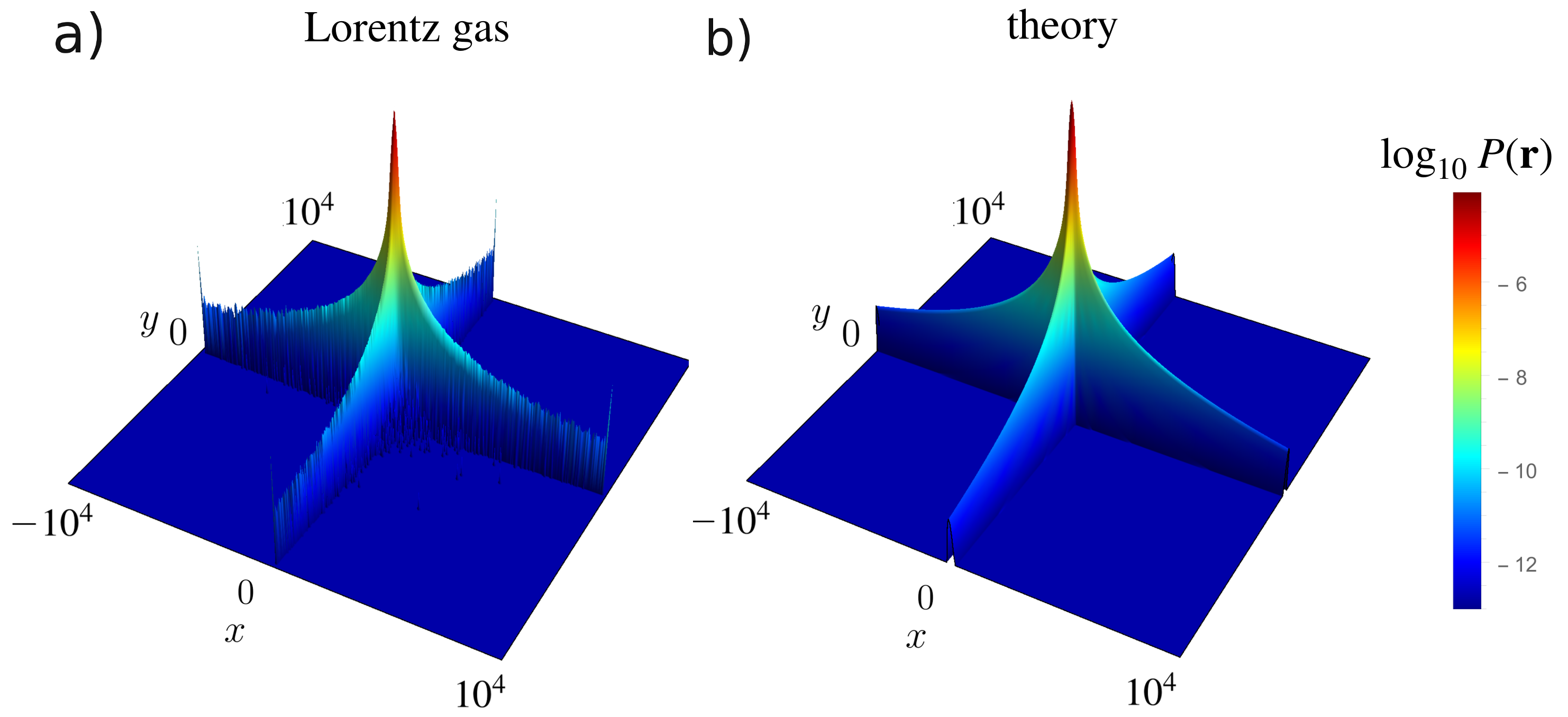}
	\caption{The probability density functions of a numerical simulation of the Lorentz gas with two open horizons (a) and the L\'evy walk (LW) theory (b) for duration $t=10^4$. The non-Gaussian crosslike shape clearly illustrates the sensitivity of the spreading density to the underlying structure of the square lattice of scatterers. The LW approximation Eq.~(\ref{eq05}) is in agreement with the simulation without any fitting. Further details can be found in the Supplemental Material \cite{SM}.}
	\label{fig2}
\end{figure*}

Using the renewal assumption, we apply the LW approach \cite{SWK,RMP,Vasily} to the LG model. We define a process where the flight times $\{\tau_n\}$ are independent identically distributed random variables drawn from the fat-tailed PDF, Eq.~(\ref{eq01}). Similarly, the velocities $\boldsymbol{v}_n$ after each collision are drawn from a PDF we denote $F(\boldsymbol{v})$. A simple geometrical calculation shows that for the chosen range of radii, one has a couple of perpendicular open horizons stretching to infinity, creating a crosslike density profile (see Fig.~\ref{fig2}). Decreasing the radius opens more horizons and results in more complex shapes (for example, when $1/\sqrt{12}<R<1/\sqrt{8}$, one finds that infinite corridors transport particles via the main diagonals as well \cite{Carl,Cristadoro}, yielding a Union Jack flag geometry). Out of this consideration, we use a velocity PDF which is aligned along the lattice's axes, and since the speed is set to one, we have $F(\boldsymbol{v}) = \{[\delta(v_x-1)+\delta(v_x+1)]\delta(v_y) + \delta(v_x)[\delta(v_y-1)+\delta(v_y+1)]\}/4$. The renewal assumption basically identifies the flight times' PDF of the LW approach with that of the LG model, and we obtain the latter.

Therefore, we devise a method to calculate the cumulative distribution function (CDF) of the walking times of the LG, $\text{CDF}(\tau)=\int_0^{\tau}\text{d}\tau'\psi(\tau')$, which was previously studied in the limit $R/a\to 0$ \cite{CDF}. The CDF is obtained from geometrical considerations and the ergodic property of the underlying process. Starting from a given scattering event, we calculate the distance to the next scatterer, and since the particles travel with unit speed this is also the time elapsed until the next collision. This time duration is controlled by two parameters, the angle of traveling direction and the initial impact parameter, which are defined in the Supplemental Material (SM) \cite{SM}. The ergodicity of the dynamics implies that the distributions of these parameters are both uniform \cite{ergodic}. We then average over all possible time durations with respect to the aforementioned parameters, thus finding an analytical expression for the traveling times' CDF. This function, which is a key ingredient for the theory presented below, exhibits rich behaviors, for example, oscillations due to the discrete nature of the scatterers' lattice arrangement [see Fig~\ref{fig1}(c)]. Asymptotically, we get the known long-time limit law of $\psi(\tau)\propto\tau^{-3}$, for which the original proof is valid in the limit $R\to a/2$ \cite{Bou85}, when the oscillations are damped out.

\begin{figure*}
	\includegraphics[width=0.75\textwidth]{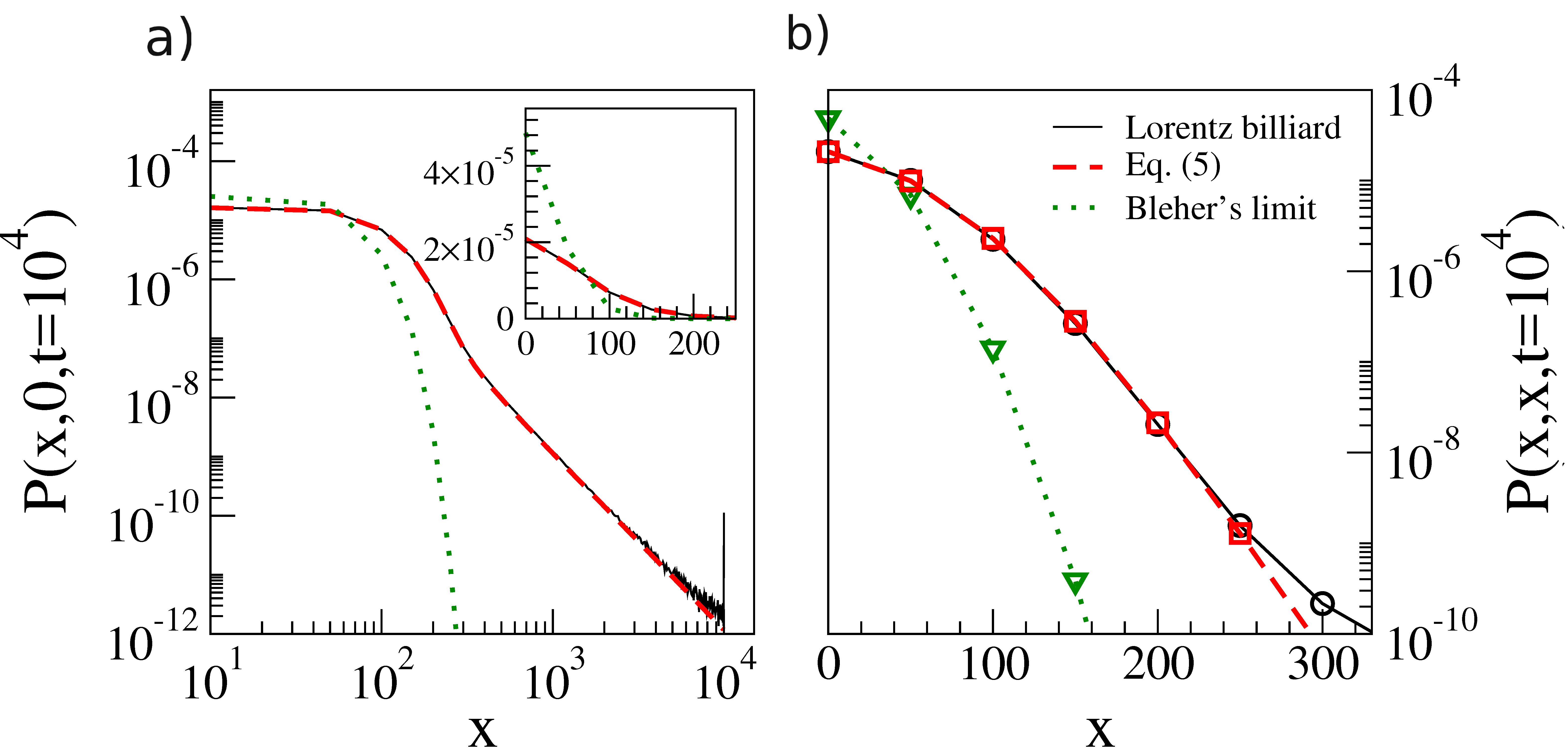}
	\caption{Cross sections of the Lorentz gas' probability density function (PDF) and the PDF given by Eq.~(\ref{eq05}) for $t=10^4$. The theory matches the simulation perfectly, both in the direction of an infinite corridor parallel to the horizontal symmetry axis (a), $y=0$, as well as in the direction of the main diagonal (b), $y=x$. The dotted green line represents Bleher's limiting law which is valid at $t\to\infty$. For (a), a linear-scaled center part is given in the inset. As this is the infinite-horizon direction, we see a power-law decay. This is in contrast with (b), where we see a fast decay with $x$, more similar to a Gaussian, due to the diagonal being blocked by scattering centers. The deviation in the last two data points of (b) originates from the finite number of sampled trajectories $\approx10^9$. Further details can be found in the Supplemental Material \cite{SM}.}
	\label{fig3}
\end{figure*}

{\em The solution.}
Let $P(\boldsymbol{r},t)$ be the density of particles, all starting at ${\bf r}(0)={\bf 0}$, and denote $\Pi(\boldsymbol{k},u)$ as its Fourier-Laplace transform, $\{\boldsymbol{r} \to \boldsymbol{k}, t \to u\}$. An exact solution for $\Pi(\boldsymbol{k},u)$ is given by the familiar Montroll-Weiss equation \cite{RMP}
\begin{equation}
\label{eq02}
	\Pi\left(\boldsymbol{k}, u\right) = \left< \frac{1 - \hat{\psi}\left(u-i\boldsymbol{k} \cdot\boldsymbol{v}\right)}{u-i\boldsymbol{k}\cdot\boldsymbol{v}} \right> \frac{1}{1 - \left< \hat{\psi} \left( u - i \boldsymbol{k} \cdot \boldsymbol{v} \right) \right> } ,
\end{equation}
where $\hat{\psi}(u)$ is the Laplace transform of $\psi(\tau)$, and the $\langle\cdots\rangle$ above denotes an average with respect to the velocity's PDF $F(\boldsymbol{v})$. To invert this equation in the long-time limit \cite{taub}, we consider the small $u$ behavior of $\hat{\psi}(u)$, derived in the SM \cite{SM},
\begin{equation}
\label{eq03}
	\hat{\psi}(u) \simeq 1 - \langle \tau \rangle u - \frac{1}{2} \left( \tau_0 u\right)^2 \ln\left( C_{\psi} \tau_0 u \right) + o\left(u^2\right) .
\end{equation}
The first term is the normalization, $\langle\tau\rangle$ is the mean time between collisions, and the last term is related to the power-law tail of $\psi(\tau)$, with $C_{\psi}$ being
\begin{align}
\label{eq04}
	C_{\psi} = \exp\left\{\vphantom{\int_{\tau_0}^{\infty} \text{d}\tau \left[ \psi(\tau)\left(\frac{\tau}{\tau_0}\right)^2 - \frac{1}{\tau}\right]} \gamma \right. & - \frac{3}{2} - \int_{0}^{\tau_0} \text{d}\tau \, \psi(\tau) \left(\frac{\tau}{\tau_0}\right)^2 \\
	&- \left. \int_{\tau_0}^{\infty} \text{d}\tau \left[ \psi(\tau)\left(\frac{\tau}{\tau_0}\right)^2 - \frac{1}{\tau}\right] \right\} \nonumber ,
\end{align}
where $\gamma \approx 0.5772$ is Euler's constant. Importantly, we obtain the parameters $\langle\tau\rangle$, $\tau_0$, and $C_{\psi}$ out of the geometrical theory of $\text{CDF}(\tau)$. The packet of spreading particles in the long-time limit is found with an asymptotic small $\{\boldsymbol{k},u\}$ expansion of Eq.~(\ref{eq02}), performed in the SM \cite{SM},
\begin{widetext}
	\begin{equation}
	\label{eq05}
		P(\boldsymbol{r},t) \simeq \frac{1}{\pi \xi^2(t)} \exp\left[ - \frac{r^2}{\xi^2(t)} \right] \left\{ 1 + \frac{1}{\Omega(t)}\sum_{j=1}^2\left\{ \left[2-\gamma-\ln \left( 4\right) \vphantom{\frac{r_j^2}{\xi^2(t)}}\right] \left[ \frac{1}{2}- \frac{r_j^2}{\xi^2(t)} \right] - \frac{1}{2} \text{M}^{(1,0,0)} \left[-1;\frac{1}{2}; \frac{r_j^2}{\xi^2(t)}\right] \right\} \right\} ,
	\end{equation}
\end{widetext}
where
\begin{align}
\label{eq06}
	&\xi(t) = \Xi\sqrt{\frac{t}{2T}\Omega(t)} , \quad && \Xi \equiv 2 C_{\psi}\tau_0V , \nonumber \\
	&\Omega(t) = \left|\text{W}_{-1}\left(-\frac{2T}{t}\right)\right| , \quad &&T\equiv 4C_{\psi}^2\left<\tau\right> ,
\end{align}
with $\boldsymbol{r}=(r_1,r_2)=(x,y)$, and, as mentioned, $V=1$. Here, $\text{M}(\cdots)$ is Kummer's confluent hypergeometric function \cite{Kummer}, and the superscript over $\text{M}$ denotes the derivative with respect to its first argument. $\text{W}_{-1}(\eta)$ is the secondary branch of the Lambert W function \cite{Kummer}, defined for $\eta\in[-1/e,0)$ by the identity $\text{W}_{-1}(\eta) = \ln[ \eta /\text{W}_{-1}(\eta)]$, which has the following expansion as $\eta\rightarrow 0^-$,
\begin{equation}
\label{eq07}
	\left|\text{W}_{-1} (\eta)\right| = L_1 + L_2 + \frac{L_2}{L_1} + \textit{O}\left(\frac{L_2^2}{L_1^2}\right) ,
\end{equation}
where $L_1=\ln(1/|\eta|)$ and $L_2=\ln[\ln(1/|\eta|)]$. As shown in Figs.~\ref{fig2} and \ref{fig3}, the solution Eq.~(\ref{eq05}) perfectly matches the simulations without any fitting, and it nicely captures the three main features of our analysis: (I) The underlying symmetry of the scatterers is reflected in the crosslike shape of the packet of particles, (II) the longstanding problem of the ultraslow convergence is solved (see below), and (III) a power-law decay of the distribution along the open horizons.

By virtue of Eqs.~(\ref{eq06}) and (\ref{eq07}), we have $\Omega(t)\simeq \ln(t)$ when $t\to\infty$, and thus the displacement $|\boldsymbol{r}|$ scales as $\sqrt{t\ln(t)}$, as was shown in Ref.~\cite{Bleher}. However, considering the correction to this leading order, $L_2$ in Eq.~(\ref{eq07}), we see that one needs to demand that $\ln(t)\gg\ln[\ln(t)]$, and as such the convergence to this mathematical limit is ultraslow. The Lambert scaling approach resolves this problem for any \textit{reasonably} large $t$, namely, time for which $\Omega(t)\gg 1$, by compactly enclosing all of the $t$-dependent logarithmic behaviors into a single function, i.e., the Lambert W function. This also means that our theory can be regarded as a series expansion in powers of \textit{a single} large parameter $\Omega(t)$, in contrast with a nested variety of logarithmic expressions which one would receive by using the standard $t\ln(t)$ scaling in a perturbative expansion, as Eq.~(\ref{eq07}) suggests. The Lambert W function provides a more accurate scaling for the Gaussian limiting form found by Bleher using the $t\ln(t)$ scaling (see the inset of Fig.~\ref{fig3}). In addition, the Kummer function's term in Eq.~(\ref{eq05}) yields for large $r_j$ the power-law behavior $P(\boldsymbol{r},t) \propto |r_j|^{-3}$. These non-Gaussian tails which decay with an exponent $-3$ are clearly related to the fat tail of the flight times PDF $\psi(\tau)\propto\tau^{-3}$. It follows that the Lambert scaling and Kummer correction found here are a required necessity for a numerical analysis, as seen in Fig.~\ref{fig3}. Equation~(\ref{eq05}) represents the packet's PDF very well, and as such one can disregard its correction $\sim\textit{O}[1/\Omega^2(t)]$.

Our solution Eq.~(\ref{eq05}) contains three parameters, i.e., $\langle\tau\rangle$, $\tau_0$, and $C_{\psi}$, all of which we are able to extract out of our geometrical theory for the CDF of the flight times [Fig.~\ref{fig1}(c)], as mentioned (see SM \cite{SM}). Furthermore, using our solution we are able to find a closed-form expression for $\tau_0$, given previous rigorous results for $\langle\tau\rangle$: Considering extremely-long-time durations for Eqs.~(\ref{eq05}) and (\ref{eq06}), namely, $t$ for which $\Omega(t)\simeq\ln(t)$, yields a Gaussian profile with a variance of $\sigma^2 t\ln(t)$, where $\sigma^2=\Xi^2/2T=\tau_0^2/2\langle\tau\rangle$. In Ref.~\cite{Bleher}, it was rigorously proven that for $t\to\infty$ the random variable $[\boldsymbol{r}(t) - \boldsymbol{r}(0)]/\sqrt{t\ln(t)}$ converges in distribution to a Gaussian variable with a zero mean and a variance $\sigma$ which is given by the scatterers' radius as
\begin{align}
\label{eq08}
	\sigma^2 = \frac{2(1-2R)^2}{\pi(1-\pi R^2)} ,
\end{align}
while for the mean time between collisions one has \cite{Carl}
\begin{align}
\label{eq09}
	\left<\tau\right> = \frac{1-\pi R^2}{2R} .
\end{align}
Thus, using the above, we find for $\tau_0$,
\begin{align}
\label{eq10}
	\tau_0 = \sqrt{2\langle\tau\rangle\sigma^2} = \sqrt{\frac{2}{\pi R}}\left(1-2R\right),
\end{align}
with an agreement to its leading behavior found in Ref.~\cite{Bou85}. Comparing Eq.~(\ref{eq10}) with the values obtained via our geometrically calculated $\psi(\tau)$ gives a good agreement for radii in the range $(1/\sqrt{8},1/2)$. Hence, the only parameter which requires the computation of $\psi(\tau)$ is $C_{\psi}$, and the tail of $\psi(\tau)$ Eq.~(\ref{eq01}) is found in closed form.

{\em Discussion and summary.}
In Ref.~\cite{Carl}, an interesting doubling effect was pointed out. While for familiar diffusion processes, e.g., Brownian motion, the Gaussian packet's variance is equal to the mean-square displacement (MSD), in this case there exists a factor of $2$ between them. It arises from the fat-tail behavior of the packet of particles, as the MSD has two contributing elements, the far tail $\sim|\boldsymbol{r}|^{-3}$ found here and the Gaussian bulk. As only half of the MSD can be explained using the Gaussian approximation, one needs to go beyond it. In this sense, the power-law tail is needed for a correct description of the MSD, which is the standard quantifier of diffusive processes. Calculation of the MSD demands the introduction of a far-tail cutoff, namely the density is zero beyond $|\boldsymbol{r}|=t$ (see Figs.~\ref{fig2} and \ref{fig3}). In order to receive a full description of the problem, one must construct a theory moving from that end point $|\boldsymbol{r}|=t$ inward, e.g., to introduce the infinite covariant density \cite{ICD}.

Our theory provides a description for the dynamics of the two-dimensional infinite-horizon LG based on the LW approach. With a correct choice of $F(\boldsymbol{v})$ and an appropriate calculation of $\psi(\tau)$, this theory can be extended to arbitrary lattice geometries, as well for other models and systems which exhibit similar features to the infinite horizon LG model \cite{kin,obs,mushroom,bac,emp}. Importantly, since the power-law behavior Eq.~(\ref{eq01}) is valid for any spatial dimension $d<6$ of the LG, so do our findings. While for the intermediate times we found smooth behavior of the particle's PDF, for short enough times one finds oscillations in $P(\boldsymbol{r},t)$ (see Fig.~1 of Ref.~\cite{Carl}). These clearly originate from the stairlike structure of $\text{CDF}(\tau)$ [see Fig.~\ref{fig1}(c)].

Finally, we have carried out numerical simulations of a one-dimensional chain of stadium billiards \cite{channel}. We found that the PDF produced by this model perfectly fits a one-dimensional variant of Eq.~(\ref{eq05}). This is further evidence that our findings are universal and irrespective of the system's spatial dimension, assuming the infinite horizon and chaotic (renewal) conditions are met.

\begin{acknowledgments}

{\em Acknowledgments.}
This work was supported by the Israel Science Foundation Grant No.~1898/17 (LZ, IF and EB). Numerical simulations were supported by the Russian Science Foundation Grant No.~16-12-10496 (SD).

\end{acknowledgments}

\setcounter{equation}{0}
\setcounter{figure}{0}
\renewcommand{\theequation}{SM\arabic{equation}}
\renewcommand{\thefigure}{SM\arabic{figure}}

\begin{widetext}

	\section{Supplemental Material}

	\subsection{Calculation of the distribution of traveling times $\psi(\tau)$}

	We begin by defining a two-dimensional square lattice of constant $a=1$ occupied with circular scatterers of radius $R$, such that the center of each circle is located on a grid point. We focus on the origin, and assume that the particle has just collided with the $(0,0)$ scatterer. We define the collision's impact parameter and recoil direction as $b$ and $\beta$ respectively, see Fig.~\ref{figSM1}. We now denote as $\tau^*(\beta,b,R)$ the time duration until the following collision, and since $V=1$ it is also the distance traveled till the next scatterer. One can then write the probability density function (PDF) $\psi(\tau)$ as:
	\begin{equation}
	\label{eqSM01}
		\psi(\tau,R) = \int_0^{2\pi} \frac{\text{d}\beta}{2\pi} \int_{-R}^{R} \frac{\text{d}b}{2R} \; \delta\left[ \tau - \tau^*(\beta,b,R) \vphantom{\frac{1}{2}} \right] ,
	\end{equation}
	where the factors of $1/2\pi \times 1/2R$ are the distributions of $\beta$ and $b$, which are both uniform due to ergodicity \cite{ergodic}. However, it is more convenient to consider each scatterer's contribution separately. Given $\beta$ and $b$ for which the particle's path ends in a specific scatterer $(n,m)$, we find the distance traveled till that scatterer:
	\begin{equation}
	\label{eqSM02}
		\tau_{n,m}^*(\beta,b,R) = n\cos(\beta) + m\sin(\beta) -\sqrt{R^2-b^2} -\sqrt{ R^2 - \left[ m\cos(\beta) - n\sin(\beta) - b \right]^2} .
	\end{equation}
	Here the integers $n$ and $m$ are the lattice coordinates of the scatterer. Eq.~(\ref{eqSM01}) then becomes:
	\begin{equation}
	\label{eqSM03}
		\psi(\tau,R) = \sum_{n,m} \int_{\beta_{n,m}^{\rm min}(R)}^{\beta_{n,m}^{\rm max}(R)} \frac{\text{d}\beta}{2\pi} \int_{b_{n,m}^{\rm min}(\beta,R)}^{b_{n,m}^{\rm max}(\beta,R)} \frac{\text{d}b}{2R} \; \delta\left[ \tau - \tau^*_{n,m}(\beta,b,R) \vphantom{\frac{1}{2}} \right] ,
	\end{equation}
	where the summation is carried over all integers besides the pair $(0,0)$, and the integration boundaries will be defined in a few steps. Assuming that $1/\sqrt{8}<R<1/2$, one has a single pair of infinite corridors, which allows us to break the problem into three parts using symmetry considerations. The first part is the four nearest neighbors' contributions, namely the scatterers $(1,0)$, $(0,1)$, $(-1,0)$ and $(0,-1)$. The second part is the four next to nearest neighbors' contributions, i.e. the circles $(1,1)$, $(-1,1)$, $(1,-1)$ and $(-1,-1)$. The third part consists out of the distant neighbors, which are the $(1,m>2)$ set of scatterers, and its eight counterparts: $(-1,m>2)$, $(1,m<-2)$, $(-1,m<-2)$, $(n>2,1)$, $(n>2,-1)$, $(n<-1,1)$ and $(n<-1,-1)$. Notice that the third part contains twice as elements with respect to each of the first/second parts. Therefore, the integration boundaries of $\beta$ for the first two parts are defined as half of their maximal value, which is possible due to symmetry, such that this factor of $2$ vanishes. We do the same for all of $b$'s boundaries, as the resulted expressions are somewhat simpler in this way. Thus, Eq.~(\ref{eqSM03}) changes to:
	\begin{equation}
	\label{eqSM04}
		\psi(\tau,R) = 16\sum_{n=0}^{\infty} \int_{\beta_{n,1}^{\rm min}(R)}^{\beta_{n,1}^{\rm max}(R)} \frac{\text{d}\beta}{2\pi} \int_{b_{n,1}^{\rm min}(\beta,R)}^{b_{n,1}^{\rm max}(\beta,R)} \frac{\text{d}b}{2R} \; \delta\left[ \tau - \tau^*_{n,1}(\beta,b,R) \vphantom{\frac{1}{2}} \right] .
	\end{equation}
	To obtain expressions for the boundaries of $\beta$, we divide them into two contributions: $\beta_{n,1}^{\rm cen}(R)$ which arises from the general direction of the $(n,1)$ scatterer, which is defined by its center, and $\beta_{n,1}^{\rm dif}(R)$, which arises from the scatterer's boundaries. We now split the $\beta$ domain of integration into two parts, as we need to take into consideration a shadowing effect, in which certain scatterers block the particle's path from reaching the destination scatterer. We denote the border between these parts as $\beta_{n,1}^{\rm turn}(R)$. Calculating the boundaries of $b$ for each of the $\beta$ domains, we find:
	\begin{equation}
	\label{eqSM05}
		b_{n,1}^{\rm min}(R) = \left\{
		\begin{aligned}
			&\cos(\beta)-n\sin(\beta)-R & \beta_{n,1}^{\rm min}(R)<\beta<\beta_{n,1}^{\rm turn}(R) \\
			&R-\sin(\beta) & \beta_{n,1}^{\rm turn}(R)<\beta<\beta_{n,1}^{\rm max}(R)
		\end{aligned}
		\right. , \quad
		b_{n,1}^{\rm max}(R) = \frac{1}{2}\left[\cos(\beta)-n\sin(\beta) \vphantom{\frac{1}{2}} \right] ,
	\end{equation}
	and:
	\begin{equation}
	\label{eqSM06}
		\beta_{n,1}^{\rm min}(R) = \beta_{n,1}^{\rm cen}(R) - \beta_{n,1}^{\rm dif}(R) , \quad \beta_{n,1}^{\rm max}(R) = \left\{
		\begin{aligned}
			&\frac{\pi}{2} & n=0 \\
			&\frac{\pi}{4} & n=1 \\
			&\beta_{n-2,1}^{\rm min}(R) & n>1
		\end{aligned}
		\right. ,
	\end{equation}
	where:
	\begin{equation}
	\label{eqSM07}
		\beta_{n,1}^{\rm cen}(R) = \left\{
		\begin{aligned}
			&\frac{\pi}{2} & n=0 \\
			&\tan^{-1}\left(\frac{1}{n}\right) & n>0
		\end{aligned}
		\right. , \quad
		\beta_{n,1}^{\rm dif}(R) = \sin^{-1}\left(\frac{2R}{n^2+1}\right) , \quad
		\beta_{n,1}^{\rm turn}(R) = \left\{
		\begin{aligned}
			&\frac{\pi}{2} & n=0 \\
			&\beta_{n-1,1}^{\rm min}(R) & n>0
		\end{aligned}
		\right. .
	\end{equation}
	Basically, one can now calculate $\psi(\tau)$ using a computational program like Mathematica and extract the needed constants out of it.

	However, specifically for the constants $\langle\tau\rangle$, $\tau_0$, and $C_{\psi}$, one can use a simpler tactic. To calculate the mean time between collisions, we simply plug Eq.~(\ref{eqSM04}) into the definition of $\langle\tau\rangle$. The Dirac delta function is then replaced with $\tau_{n,m}^*(\beta,b,R)$, and to achieve the designated precision one can simply truncate the sum at a large enough $M$. For $M=500$ we obtained $\langle\tau\rangle\approx 0.62153$, with a relative error of $0.025\%$ to the analytical result Eq.~(9). The constant $\tau_0$, which is defined as $\lim_{\tau\to\infty}\tau^3\psi(\tau)=\tau_0^2$ can be dealt with in a similar way. It follows out of L'Hospital's rule that:
	\begin{equation}
	\label{eqSM08}
		\tau_0^2 = \lim_{T\to\infty} \frac{1}{T} \int_{0}^{T} \text{d}\tau \, \tau^3 \, \psi(\tau) .
	\end{equation}
	Plugging Eq.~(\ref{eqSM04}) into Eq.~(\ref{eqSM08}) and performing the integral over $\tau$, the Dirac delta function becomes a Heaviside step function. This in turn truncates the sum in Eq.~(\ref{eqSM04}) at a certain $M$. It is easy to show that $\tau_{M,1}^*(\beta,b,R)\simeq M$ for large integer $M$. Thus, we change $T\to M$, obtaining:
	\begin{equation}
	\label{eqSM09}
		\tau_0^2 = \lim_{M\to\infty} \frac{16}{M}\sum_{n=0}^M \int_{\beta_{n,1}^{\rm min}(R)}^{\beta_{n,1}^{\rm max}(R)} \frac{\text{d}\beta}{2\pi} \int_{b_{n,1}^{\rm min}(\beta,R)}^{\phi_{n,1}^{\rm max}(\beta,R)} \frac{\text{d}\phi}{2R} \; \tau^{*3}_{n,1}(\beta,b,R) .
	\end{equation}
	Eq.~(\ref{eqSM09}) converges rather slowly. Since its summands are all positive, it can be changed into:
	\begin{equation}
	\label{eqSM10}
		\tau_0^2 = \lim_{M\to\infty} 16 \int_{\beta_{M,1}^{\rm min}(R)}^{\beta_{M,1}^{\rm max}(R)} \frac{\text{d}\beta}{2\pi} \int_{b_{M,1}^{\rm min}(\beta,R)}^{\phi_{M,1}^{\rm max}(\beta,R)} \frac{\text{d}\phi}{2R} \; \tau^{*3}_{M,1}(\beta,b,R) .
	\end{equation}
	For $M=5000$, we obtained $\tau_0\approx 0.25238$, with a relative error of $0.028\%$ to our analytical result Eq.~(10). Finally, for $C_{\psi}$ we obtain the following formula out of Eq.~(4):
	\begin{equation}
	\label{eqSM11}
		C_{\psi} = \lim_{T\to\infty} \exp\left[ \gamma - \frac{3}{2} - \int_0^T\text{d}\tau \, \psi(\tau) \left( \frac{\tau}{\tau_0} \right)^2 + \ln\left( \frac{\tau}{\tau_0} \right) \right] .
	\end{equation}
	Combining it with Eq.~(\ref{eqSM04}) yields:
	\begin{equation}
	\label{eqSM12}
		C_{\psi} = \lim_{M\to\infty} \exp\left[ \gamma - \frac{3}{2} + \ln\left( \frac{M}{\tau_0} \right) - 16\sum_{n=0}^M \int_{\beta_{n,1}^{\rm min}(R)}^{\beta_{n,1}^{\rm max}(R)} \frac{\text{d}\beta}{2\pi} \int_{b_{n,1}^{\rm min}(\beta,R)}^{\phi_{n,1}^{\rm max}(\beta,R)} \frac{\text{d}\phi}{2R} \frac{\tau^{*2}_{n,1}(\beta,b,R)}{\tau_0^2} \right] ,
	\end{equation}
	and we receive for $M=5000$ the value $C_{\psi}\approx 4.4816\cdot 10^{-4}$.

	\subsection{Leading behavior of $\hat{\psi}(u \rightarrow 0)$}

	Here we derive Eq.~(3), obtaining Eq.~(4) during the process. We assume that $\psi(\tau)$ behaves asymptotically as
	\begin{equation}
	\label{eqSM13}
		\lim_{\tau \rightarrow \infty} \psi(\tau)\tau^3 = \tau_0^2 , \quad \tau_0>0 ,
	\end{equation}
	with its Laplace transform being defined by
	\begin{equation}
	\label{eqSM14}
		\hat{\psi}(u) = \int_{0}^{\infty} \text{d}\tau \, \psi(\tau) e^{-u\tau} .
	\end{equation}
	We will now show that using Eq.~(\ref{eqSM13}), Eq.~(\ref{eqSM14}) reduces to Eq.~(3) for small $u$. To do so, we rewrite Eq.~(\ref{eqSM14}) as:
	\begin{equation}
	\label{eqSM15}
		\hat{\psi}(u) = 1 - u\left<\tau\right> + \int_{0}^{\infty} \text{d}\tau \, \psi(\tau) \left[ e^{-u\tau} - 1 + u\tau \vphantom{\frac{1}{2}}\right] .
	\end{equation}
	In order to find the leading behavior of the last term of Eq.~(\ref{eqSM15}), the following limit is considered:
	\begin{equation}
	\label{eqSM16}
		l_1=\lim_{u \rightarrow 0} \frac{1}{u^2\ln(u)} \int_{0}^{\infty} \text{d}\tau \, \psi(\tau) \left[ e^{-u\tau} - 1 + u\tau \vphantom{\frac{1}{2}}\right] .
	\end{equation}
	Using L'Hospital's rule three times, followed by a change of variable to $\eta=u\tau$, Eq.~(\ref{eqSM16}) takes the form:
	\begin{equation}
	\label{eqSM17}
		l_1=-\frac{1}{2} \lim_{u \rightarrow 0} \int_{0}^{\infty} \text{d}\eta \, \psi\left(\frac{\eta}{u}\right) \left(\frac{\eta}{u}\right)^3 e^{-\eta} .
	\end{equation}
	For any finite $u$, the integrand vanishes for $\eta=0$, as $\psi(\tau)$ is normalized. Since a finite set of bounded points does not contribute to an integral, we may discard this point, such that the integration is carried over the domain $(0,\infty)$. This allows us to employ the dominated convergence theorem, and switch the order of limit and integration in Eq.~(\ref{eqSM17}). Together with Eq.~(\ref{eqSM13}), we have:
	\begin{equation}
	\label{eqSM18}
		l_1=-\frac{\tau_0^2}{2} \int_{0^{+}}^{\infty} \text{d}\eta \, e^{-\eta} = -\frac{\tau_0^2}{2} ,
	\end{equation}
	which shows that the limit $l_1=-\tau_0^2/2$ is finite, yielding:
	\begin{equation}
	\label{eqSM19}
		\hat{\psi}(u) \simeq 1 - \langle \tau \rangle u +l_1 u^2 \ln(u) + \mathcal{O} \left( u^2 \right) .
	\end{equation}
	To complete the derivation, we calculate the next order correction $\sim u^2$. For that cause, the following limit is considered:
	\begin{equation}
	\label{eqSM20}
		l_2 = \lim_{u \rightarrow 0} \frac{1}{u^2} \left\{ \int_{0}^{\infty} \psi(\tau) \left[ e^{-u\tau} - 1 + u\left<\tau\right> \vphantom{\frac{1}{2}} \right] \text{d}\tau + \frac{\tau_0^2}{2} u^2\ln(u) \right\} .
	\end{equation}
	Using L'Hospital's rule two times, followed by a split of the integral at $\tau=\tau_0$, Eq.~(\ref{eqSM20}) becomes:
	\begin{equation}
	\label{eqSM21}
		l_2=\frac{1}{2}\lim_{u \rightarrow 0} \int_{0}^{\tau_0} \text{d}\tau \, \psi(\tau) \tau^2 e^{-u\tau} + \frac{1}{2}\lim_{u \rightarrow 0} \left\{ \int_{\tau_0}^{\infty} \text{d}\tau \psi(\tau) \tau^2 e^{-u\tau} + \tau_0^2 \left[ \ln(u) + \frac{3}{2} \right] \right\} .
	\end{equation}
	The first integral of Eq.~(\ref{eqSM21}) is carried over a finite region, so one can exchange the order of limit and integration. Moving to the second integral, we add and subtract the term $\tau_0^2/\tau^3$ from $\psi(\tau)$, which yields:
	\begin{equation}
	\label{eqSM22}
		l_2 = \frac{1}{2} \int_{0}^{\tau_0} \text{d}\tau \, \psi(\tau) \tau^2 + \frac{1}{2}\lim_{u \rightarrow 0} \int_{\tau_0}^{\infty} \text{d}\tau \left[\psi(\tau)-\frac{\tau_0^2}{\tau^3}\right] \tau^2 e^{-u\tau} + \frac{1}{2}\lim_{u \rightarrow 0} \left\{ \int_{\tau_0}^{\infty} \text{d}\tau \frac{\tau_0^2}{\tau} e^{-u\tau} + \tau_0^2 \left[ \ln(u) + \frac{3}{2} \right] \right\} .
	\end{equation}
	Note that due to the asymptotics Eq.~(\ref{eqSM13}), the second integral in Eq.~(\ref{eqSM22}) converges when $u \rightarrow 0$. Evaluating the middle row limit and the bottom row integral, we obtain:
	\begin{equation}
	\label{eqSM23}
		l_2 = \frac{1}{2} \int_{0}^{\tau_0} \text{d}\tau \, \psi(\tau) \tau^2 +\frac{1}{2} \int_{\tau_0}^{\infty} \text{d}\tau \left[\psi(\tau)-\frac{\tau_0^2}{\tau^3}\right] \tau^2 + \frac{1}{2}\lim_{u \rightarrow 0} \tau_0^2 \left[ \Gamma\left(0,\tau_0u\right) + \ln(u) + \frac{3}{2} \right] ,
	\end{equation}
	where $\Gamma(\cdot,\cdot)$ is the incomplete Gamma function, which has the small $\eta$ behavior $\Gamma(0,\eta) \simeq -\ln(\eta) -\gamma +\eta$. Thus, after some algebra one finds for the limit $l_2$:
	\begin{equation}
	\label{eqSM24}
		l_2=-\frac{\tau_0^2}{2} \ln\left(C_{\psi}\tau_0\right) ,
	\end{equation}
	where the constant $C_{\psi}$ is defined by Eq.~(4). Equation~(3) then follows from:
	\begin{equation}
	\label{eqSM25}
		\hat{\psi}(u) \simeq 1 - \langle \tau \rangle u +l_1 u^2 \ln(u) + l_2u^2 + o\left(u^2\right) .
	\end{equation}

	\subsection{Approximation of the L\'evy walk model with Lambert scaling}

	The two-dimensional L\'evy walk model is defined as follows: A random walker is placed at $\boldsymbol{r}(0)$ on time $t=0$. Its movement consists of segments of ballistic motion with constant velocity, separated by collision-like events which induce a change in the velocity's magnitude and/or direction. The process lasts for a predetermined time duration (the measurement time $t$). The model employs two PDFs in order to determine the displacement during each of the ballistic motion epochs. The velocity of each segment is drawn from a PDF $F(\boldsymbol{v})$, whose moments are all finite, and is further assumed to be symmetric with respect to each of the components $v_j$, with $j=1,2$, such that its odd moments vanish. The time duration of each ballistic section is drawn from a PDF $\psi(\tau)$. The movement continues until the allotted measurement time is met, thus the number of collisions $N$ in $(0,t)$ is random. This yields the total displacement as:
	\begin{equation}
	\label{eqSM26}
		\boldsymbol{r}(t) - \boldsymbol{r}(0) = \sum_{n=1}^{N} \boldsymbol{v}_{n-1}\tau_n + \boldsymbol{v}_{N}\tau^* ,
	\end{equation}
	with $0\le\tau^*=t-\sum_{n=1}^{N}\tau_n$. The traveling times and velocities $\{\tau_n,\boldsymbol{v}_n\}$, for $1\le n\le N$, are independent identically distributed random variables, and the initial conditions $\boldsymbol{r}(0)$ and $\boldsymbol{v}_0$ are drawn from equilibrium. Let us denote the probability to find the walker at a position $\boldsymbol{r}$ on time $t$ as $P(\boldsymbol{r},t)$. Applying Fourier and Laplace transforms to the spatial and temporal coordinates of $P(\boldsymbol{r},t)$ respectively, an exact expression of the probability in Fourier-Laplace space, denoted $\Pi(\boldsymbol{k},u)$, is given by the Montroll-Weiss Eq.~(2), where
	\begin{equation}
	\label{eqSM27}
		\left<\cdots\right> = \int_{-\infty}^{\infty} \text{d}^2v F(\boldsymbol{v}) \cdots
	\end{equation}
	(not to be confused with $\langle\tau\rangle$, which is simply the mean time between collisions). As mentioned in the main text, in order to find the density of particles while accounting for the problem's spatial structure, we take for the Lorentz gas a ``cross" velocity PDF, which represents the underlying square lattice:
	\begin{equation}
	\label{eqSM28}
		F\left(\boldsymbol{v}\right) = \frac{1}{4} \left\{ \left[ \delta\left(v_x-V\right) + \delta\left(v_x+V\right) \vphantom{\frac{v_{\times}}{\sqrt{2}}} \right] \delta\left(v_y\right) + \delta\left(v_x\right) \left[ \delta\left(v_y-V\right) + \delta\left(v_y+V\right) \vphantom{\frac{v_{\times}}{\sqrt{2}}} \right] \right\} ,
	\end{equation}
	where $V>0$ is a constant (in the main text $V=1$). This yields for the denominator of Eq.~(2):
	\begin{equation}
	\label{eqSM29}
		1 - \left<\hat{\psi}(u-i\boldsymbol{k}\cdot\boldsymbol{v})\right> = 1 - \frac{1}{4}\sum_{j=1}^2\left[\hat{\psi}\left(u-iVk_j\right) + \hat{\psi}\left(u+iVk_j\right) \right] .
	\end{equation}
	In order to use the asymptotic form Eq.~(3), we assume a scaling of $u \sim k^2L(|\boldsymbol{k}|)$, where $L(\cdot)$ is some logarithmic-behaving function. This suggests that $u \ll V|\boldsymbol{k}|$ when $|\boldsymbol{k}| \rightarrow 0$. Using the identity
	\begin{equation}
	\label{eqSM30}
		\ln\left(a\pm ib\right) = \frac{1}{2}\ln\left(a^2+b^2\right) \pm i\arctan\left(\frac{b}{a}\right)
	\end{equation}
	together with Eq.~(3), while neglecting the appropriate terms according to the above scaling assumption, one has
	\begin{equation}
	\label{eqSM31}
		\hat{\psi}\left(u\pm iVk_j\right) \simeq 1 - \left<\tau\right>\left(u\pm iVk_j\right) - \frac{1}{4}\left(\tau_0Vk_j\right)^2 \ln\left[\left(C_{\psi}\tau_0Vk_j\right)^2\right] \pm i\frac{\pi}{4} \left(\tau_0Vk_j\right)^2 .
	\end{equation}
	Plugging this into Eq.~(\ref{eqSM29}) while discarding irrelevant terms with respect to the above scaling assumption, we get for the denominator of Eq.~(2):
	\begin{equation}
	\label{eqSM32}
		1 - \left<\hat{\psi}(u-i\boldsymbol{k}\cdot\boldsymbol{v})\right> \simeq u\left<\tau\right> - \frac{1}{8} \sum_{j=1}^2 \left(\tau_0Vk_j\right)^2 \ln\left[\left(C_{\psi}\tau_0Vk_j\right)^2\right] ,
	\end{equation}
	while for the numerator we have:
	\begin{equation}
	\label{eqSM33}
		\left<\frac{1 - \hat{\psi}(u-i\boldsymbol{k}\cdot\boldsymbol{v})}{u-i\boldsymbol{k}\cdot\boldsymbol{v}}\right> \simeq \left< \tau \right> .
	\end{equation}
	Therefore we find, using Eq.~(2):
	\begin{equation}
	\label{eqSM34}
		\Pi(\boldsymbol{k},u) \simeq \left[u - \frac{\tau_0^2V^2}{8\left<\tau\right>} \sum_{j=1}^2 k_j^2 \ln\left( C_{\psi}^2 \tau_0^2 V^2 k_j^2 \right) \right]^{-1} .
	\end{equation}
	Returning to the time domain yields for the Fourier transform of $P(\boldsymbol{r},t)$:
	\begin{equation}
	\label{eqSM35}
		\tilde{P}(\boldsymbol{k},t) \simeq \exp\left[ t\frac{\tau_0^2V^2}{8\left<\tau\right>} \sum_{j=1}^2 k_j^2 \ln\left( C_{\psi}^2 \tau_0^2 V^2 k_j^2 \right) \right] .
	\end{equation}
	Defining $\boldsymbol{\kappa}=\boldsymbol{k} \sqrt{\tau_0^2V^2t\Omega(t)/8\left<\tau\right>}$, where $\Omega(t)$ is a yet unknown scaling function, leads to:
	\begin{equation}
	\label{eqSM36}
		\tilde{P}(\boldsymbol{\kappa},t) \simeq \frac{8\left<\tau\right>}{\tau_0^2V^2t\Omega(t)} \exp\left\{ \sum_{j=1}^2 \frac{\kappa_j^2}{\Omega(t)} \ln\left[ \frac{8 C_{\psi}^2 \left<\tau\right>}{t\Omega(t)} \kappa_j^2 \right] \right\} .
	\end{equation}
	Demanding that $\ln[t\Omega(t)/(8C_{\psi}^2\langle\tau\rangle)]=\Omega(t)$ yields:
	\begin{equation}
	\label{eqSM37}
		\Omega(t) = \left| \text{W}_{-1}\left( - 8C_{\psi}^2 \frac{\left<\tau\right>}{t} \right) \right| ,
	\end{equation}
	thus we obtain the following form for (\ref{eqSM36}):
	\begin{equation}
	\label{eqSM38}
		\tilde{P}(\boldsymbol{\kappa},t) \simeq \frac{8\left<\tau\right>e^{-\kappa^2}}{\tau_0^2V^2t\Omega(t)} \exp\left[ \sum_{j=1}^2 \frac{\kappa_j^2}{\Omega(t)} \ln\left(\kappa_j^2\right) \right] \simeq \frac{8\left<\tau\right>e^{-\kappa^2}}{\tau_0^2V^2t\Omega(t)} \left[1 + \sum_{j=1}^2 \frac{\kappa_j^2}{\Omega(t)} \ln\left(\kappa_j^2\right) \right] .
	\end{equation}
	The assumption of large $t$ constricts us to large $\Omega(t)$, and thus the second exponential term of Eq.~(\ref{eqSM38}) can be expanded to sub-leading order. The inverse Fourier transform is:
	\begin{equation}
	\label{eqSM39}
		P(\boldsymbol{r},t) = \int \frac{\text{d}^2\kappa}{(2\pi)^2} \tilde{P}\left(\boldsymbol{\kappa},t\right) \cos\left[ \frac{ \boldsymbol{\kappa} \cdot \boldsymbol{r} }{\tau_0V} \sqrt{\frac{8\left<\tau\right>}{ t \Omega(t)} } \right] .
	\end{equation}
	Evaluating these integrals, the leading and sub-leading orders of Eq.~(\ref{eqSM39}) result in Eqs.~(5) and (6). The mean square displacement (MSD) can be calculated from differentiating the Montroll-Weiss Eq.~(2). We find that in the long time limit:
	\begin{equation}
	\label{eqSM40}
		\left<r^2(t)\right> \simeq \tau_0^2V^2 \frac{t}{\left<\tau\right>} \left[ \ln\left( \frac{t}{C_{\psi}\tau_0} \right) - 2 + \gamma \right] .
	\end{equation}
	Using Eq.~(6), in the extremely long time limit we have $\text{MSD}\simeq2\sigma^2t\ln(t)$, with $\sigma^2=\tau_0^2/2\langle\tau\rangle$, and thus the doubling effect can be seen.

	\subsection{Remarks about figures 2 and 3}

	In order to plot the simulations's histogram and the PDF $P(\boldsymbol{r},t)$ on the same figure, we first notice that Eq.~(5) is analytically integrable, with the following primitive function:
	\begin{align}
	\label{eqSM41}
		&\mathcal{P}(\boldsymbol{r},t) = \int_0^{y} \text{d}y' \int_0^{x} \text{d}x' P(\boldsymbol{r}',t) \simeq \\
		&\frac{1}{4} \text{erf}\left[ \frac{x}{\xi(t)} \right] \text{erf}\left[ \frac{y}{\xi(t)} \right] \left\{ 1 + \frac{1}{\sqrt{\pi}\Omega(t)} \sum_{j=1}^2 \frac{r_j}{\xi(t)} \frac{\exp\left[-r_j^2/\xi^2(t)\right]} {\text{erf}\left[r_j/\xi(t)\right]} \left\{ 2-\gamma -\ln (4) - \text{M}^{(1,0,0)}\left[ 0;\frac{3}{2};\frac{r_j^2}{\xi^2(t)} \right] \right\} \right\} \nonumber .
	\end{align}
	As mentioned in the main text, the simulation which is presented in Figs.~2 and 3 is of duration $t=10^4$, has $\approx 10^9$ sampled trajectories, a billiards radius of $R=0.4$, a lattice constant of $a=1$, and a speed $V=1$. We define the simulation's bin indexes as $n=\text{int}(x/50)$ and $m=\text{int}(y/50)$, where $\text{int}(\eta)$ is the integer part of $\eta$. Notice that in this case the $(0,0)$ bin has twice the area of an $(n>0,0)$ bin and four times the area of an $(n>0,m>0)$ bin, therefore the appropriate histogram values were divided by a relevant factor. We now define the following function, which is an analytical representation of the numerical histogram:
	\begin{align}
	\label{eqSM42}
		&\mathcal{P}_{\rm bin}(n,m,t) = \int_{50n}^{50n+50}\text{d}x \int_{50m}^{50m+50}\text{d}y P(x,y,t) = \\
		&\mathcal{P}(50n+50,50m+50,t) + \mathcal{P}(50n,50m,t) - \mathcal{P}(50n+50,50m,t) - \mathcal{P}(50n,50m+50,t) \nonumber .
	\end{align}
	This function provides us with all/half/quarter of the probability to find a particle in the $(n>0,m>0)$/$(n>0,0)$/$(0,0)$ bins, respectively. Using the values which we obtained for $\langle\tau\rangle$, $\tau_0$ and $C_{\psi}$, we calculated the values of $\mathcal{P}_{\rm bin}(n,m,t)$. Figures 2 and 3 then follow from the rescaling $P(\boldsymbol{r},t)=P(x,y,t)\approx(1/50^2)\mathcal{P}_{\rm bin}(50n,50m,t)$, where for Fig.~3 we have $m=0$ (a) and $m=n$ (b). This rescaling was also performed on the numerical histogram. Finally, the simulation's mean time between collisions, $\langle\tau\rangle_{\rm S} \approx 0.6213$, was found to be consistent with the analytical expressions.

	\begin{figure}
		\includegraphics[width=0.4\textwidth]{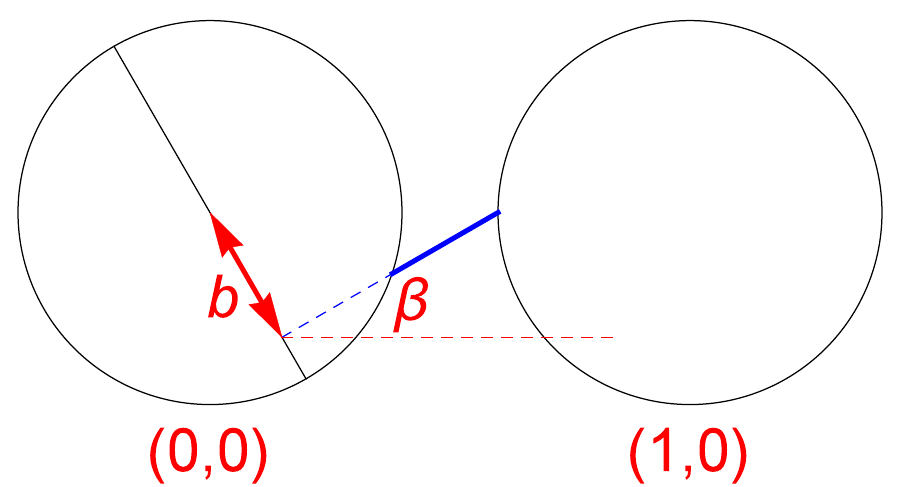}
		\caption{(color online) A sketch demonstrating the parameters which determine the trajectory's length. For this example we used $n=1$, $m=0$, $\beta=\pi/6$, $b=-0.3$ and $R=0.4$, which correspond to $\tau_{n,m}^*(\beta,b,R)\approx 0.255$ (the thick blue line between the scatterers).}
		\label{figSM1}
	\end{figure}

\end{widetext}


\begin{thebibliography}{99}

\bibitem{Bunimovich}
L. A. Bunimovich, {\em Zh. Eksp. Teor. Fiz.} {\bf 89}, 1452 (1985).

\bibitem{BSC}
L. A. Bunimovich, Ya. G. Sinai, and N. I. Chernov, {\em Russ. Math. Surv.} {\bf 45}, 105 (1990).

\bibitem{Dav}
N. Friedman, A. Kaplan, D. Carasso, and N. Davidson, {\em Phys. Rev. Lett.} {\bf 86}, 1518 (2001).

\bibitem{Ott}
D. Armstead, B. R. Hunt, and E. Ott, {\em Phys. Rev. Lett.} {\bf 89}, 284101 (2002).

%
\bibitem{Armstead}
D. N. Armstead, B. R. Hunt, and E. Ott, {\em Phys. Rev. E.} {\bf 67}, 021110 (2003).

%
\bibitem{Artuso}
R. Artuso and G. Cristadoro, {\em Phys. Rev. Lett.} {\bf 90}, 244101 (2003).

%
\bibitem{channel}
G. M. Zaslavsky and M. A. Edelman, {\em Physica D} {\bf 193}, 128-147 (2004).

%
\bibitem{Balint}
P. B\'alint and S. Go\"uzel, {\em Comm. Math. Phys.} {\bf 263}, 461 (2006).

%
\bibitem{Sanders}
D. P. Sanders and H. Larralde, {\em Phys. Rev. E}, {\bf 73}, 026205 (2006).

%
\bibitem{Szas}
D. Sa\'asz and T. Varj\'u, {\em J. Stat. Phys.} {\bf 129}, 59 (2007).

\bibitem{Courbage}
M. Courbage, M. Edelman, S. M. Saberi Fathi, and G. M. Zaslavsky, {\em Phys. Rev. E.} {\bf 77}, 036203 (2008).

%
\bibitem{Chernov}
D. I. Dolgopyat, and N. L. Chernov, {\em Russ. Math. Surv.} {\bf 64}, 651 (2009).

\bibitem{Burioni}
R. Burioni, L. Caniparoli, and A. Vezzani, {\em Phys. Rev. E.} {\bf 81}, 060101(R) (2010).

%
\bibitem{MZ}
G. Cristadoro, T. Gilbert, M. Lenci, and D. P. Sanders, {\em Phys. Rev. E} {\bf 90}, 050102 (2014).

%
\bibitem{dett}
C. P. Dettmann, {\em Commun. Theor. Phys.} {\bf 62}, 521 (2014).

\bibitem{Bianchi}
A. Bianchi, G. Cristadoro, M. Lenci, and M. Ligab\'o, {\em J. Stat. Phys.} {\bf 163}, 22 (2016).

\bibitem{Fran}
M. Spanner, F. H\'ofling, S. C. Kapfer, K. R. Mecke, G. E. Schr\"oder-Turk, and T. Franosch, {\em Phys. Rev. Lett.} {\bf 116}, 060601 (2016).

\bibitem{Carl}
C. P. Dettmann, {\em J. Stat. Phys.} {\bf 146}, 181 (2012).

%
\bibitem{Bleher}
P. M. Bleher, {\em J. Stat. Phys.} {\bf 66}, 315 (1992).

\bibitem{Cristadoro}
G. Cristadoro, T. Gilbert, M. Lenci, and D. P. Sanders, {\em Phys. Rev. E} {\bf 90}, 022106 (2014).

%
\bibitem{Bou85}
J. P. Bouchaud and P. Le Doussal, {\em J. Stat. Phys.} {\bf 41}, 225 (1985).

%
\bibitem{Zacharel}
A. Zacharl, T. Geisel, J. Nierwetberg, and G. Radons, {\em Phys. Lett. A} {\bf 114}, 317 (1986).

\bibitem{SWK}
M. F. Shlesinger, B. J. West, and J. Klafter, {\em Phys. Rev. Lett.} {\bf 58}, 1100 (1987).

\bibitem{RMP}
V. Zaburdaev, S. Denisov, and J. Klafter, {\em Rev. Mod. Phys.} {\bf 87}, 483 (2015).

\bibitem{Vasily}
V. Zaburdaev, I. Fouxon, S. Denisov, and E. Barkai, {\em Phys. Rev. Lett.} {\bf 117}, 270601 (2016).

%
\bibitem{CDF}
J. Bourgain, F. Golse, and B. Wennberg, {\em Commun. Math. Phys.} {\bf 190}, 491 (1998).

%
\bibitem{SM}
See Supplemental Material above for (I) an outline of the calculation of $\psi(\tau)$, (II) a derivation of Eq.~(\ref{eq03}), (III) a derivation of Eq.~(\ref{eq05}), and (IV) remarks about Figs.~\ref{fig2} and \ref{fig3}.

%
\bibitem{ergodic}
C. Boldrighini, L. A. Bunimovich, and Y. G. Sinai, {\em J. Stat. Phys.} {\bf 32}, 477 (1983).

%
\bibitem{taub}
J. Klafter and I. M. Sokolov, {\em First Steps in Random Walks}, Oxford University Press, New York (2011).

%
\bibitem{Kummer}
See http://functions.wolfram.com for more information.

%
\bibitem{ICD}
A. Rebenshtok, S. Denisov, P. H\"anggi, and E. Barkai, {\em Phys. Rev. Lett.} {\bf 112}, 110601 (2014).

\bibitem{mushroom}
E. G. Altmann, A. E. Motter, and H. Kantz, {\em Phys. Rev. E} {\bf 73}, 026207 (2006).

\bibitem{emp}
A. Clauser, C. R. Shalizi, and M. E. J. Newman, {\em SIAM Review} {\bf 51}, 661 (2009).

\bibitem{kin}
D. A. Kessler and E. Barkai, {\em Phys. Rev. Lett.} {\bf 108}, 230602 (2012).

\bibitem{obs}
Y. Sagi, M. Brook, I. Almog, and N. Davidson, {\em Phys. Rev. Lett.} {\bf 108}, 093002 (2012).

\bibitem{bac}
G. Ariel, A. Be’er, and A. Reynolds, {\em Phys. Rev. Lett.} {\bf 118}, 228102 (2017).

\end{thebibliography}
\end{document}